\documentclass[aps,pra,twocolumn,showpacs,groupedaddress]{revtex4}
\usepackage{epsfig,amsbsy,amsmath,amssymb,graphicx,color,epstopdf}

\begin{document}

\title{A Monte Carlo simulation study on the wetting behavior of water on graphite surface}

\author{Xiongce Zhao}
\email{xiongce.zhao@nih.gov}
\affiliation{Joint Institute for Computational Sciences and Center for Nanophase Materials Sciences, Oak Ridge National Laboratory, 
Oak Ridge, Tennessee 37831, USA}
\altaffiliation{Current address: NIDDK, 
National Institutes of Health, Bethesda, MD 20892, USA}

%\date{\today}

\begin{abstract}
This paper is an expanded edition of the rapid communication published several years ago by the author (\emph{Phys. Rev. B, v76, 041402(R), 2007}) on the simulation of wetting transition of water on graphite, aiming to provide more details on the methodology, parameters, and results of the study which might be of interest to certain readers.
 We calculate adsorption isotherms
of water on graphite using grand canonical Monte Carlo simulations
combined with multiple histogram reweighting, 
based on the empirical potentials 
of SPC/E for water, the 10-4-3 van der Waals model, and a recently developed 
induction and multipolar potential for water and graphite. Our results show
that wetting transition of 
water on graphite occurs at 475-480 K, and the prewetting critical 
temperature lies in the range of 505-510 K. 
The calculated wetting transition 
temperature agrees quantitatively with a previously predicted value using a 
simple model.
 The observation of the coexistence of 
stable and metastable states at temperatures between the wetting transition
temperature and prewetting critical temperature indicates that the transition
is first order. 
\end{abstract}


\maketitle

\section{INTRODUCTION}
\begin{figure}

 \includegraphics[width=75mm]{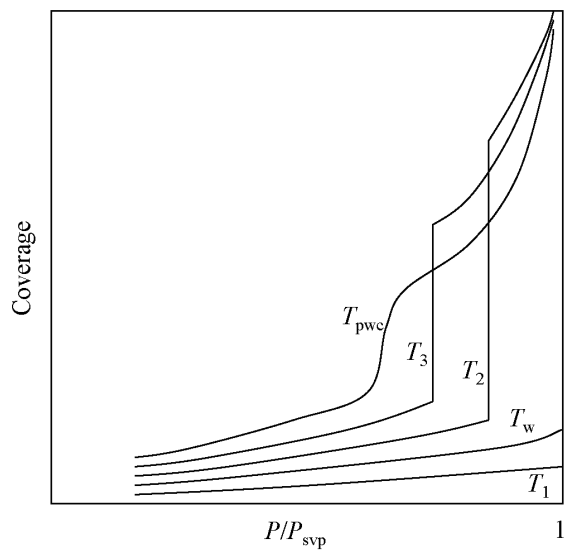}%
 \caption{Schematic diagram of adsorption isotherms near a wetting transition. 
$T_{\mathrm{pwc}}>T_3>T_2>T_{\mathrm{w}}>T_1$. \label{diagram}}

 \end{figure}

When a fluid adsorbs on a solid surface at temperatures below its 
liquid-vapor critical temperature ($T_{\rm c}$), the adsorbed film either 
spreads across the surface (wetting) or beads up as a droplet (nonwetting) 
as the pressure approaches the saturated vapor pressure \textit{P}$_{\mathrm{svp}}$ 
of the fluid. Wetting transition describes the transition between those two 
kinds of behavior.  Physically the wetting transition corresponds to the 
phenomena when the contact angle of the liquid drop on the surface changes 
from a nonzero value to zero. Analysis of wetting transition was first 
presented 30 years ago by Cahn \cite{Cahn:1977} and Ebner and 
Saam. \cite{Ebner:1977} They showed that if a fluid does not wet a 
particular surface at low temperature, then the system ought to exhibit wetting transition 
at some temperature \textit{T}$_{\mathrm{w}}$ below $T_{\rm c}$.  
In terms of adsorption isotherms, the 
wetting phenomenon should manifest itself as following three different 
patterns. (1) At temperatures below \textit{T}$_{\mathrm{w}}$, adsorption beginning 
with a thin film increases slightly as the pressure increases towards the saturation pressure 
\textit{P}$_{\mathrm{svp}}$. At \textit{P}$_{\mathrm{svp}}$ the bulk vapor 
condenses completely, and the adsorption coverage becomes infinite. On a 
coverage versus pressure diagram, the adsorption isotherm reaches 
\textit{P}$_{\mathrm{svp}}$ with a discontinuous jump (infinite slope). (2) In the 
temperature range between \textit{T}$_{\mathrm{w}}$ and the prewetting critical 
temperature \textit{T}$_{\mathrm{pwc}}$, the thin film grows as the pressure increases 
until it jumps to a thick, liquid like film of finite thickness at some 
pressure less than \textit{P}$_{\mathrm{svp}}$. This thin-to-finite film transition 
 or wetting transition
is followed by continuous growth until condensation occurs at 
\textit{P}$_{\mathrm{svp}}$. (3) At temperatures higher than \textit{T}$_{\mathrm{pwc}}$, the 
film grows continuously and the prewetting transition 
disappears.  A schematic diagram showing these three types of adsorption 
patterns is given in FIG. \ref{diagram}. 

Since the first theory on wetting transition was developed a variety of 
experimental \cite{Mistura:1994,Ross:1995,Hallock:1995,Demolder:1995,
Wyatt:1995,Yao:1996,Hess:1997,Ross:1997,Kozhevnikov:1997,Ross:1998,
Hensel:1998,Ohmasa:1998,Ohmasa:2001,Kozhevnikov:1998} 
and theoretical studies 
\cite{Ebner:1987,Finn:1989,Cheng:1991,Cheng:1993,Wagner:1994,
Bojan:1998,Boninsegni:1998,Ancilotto:1999,Bojan:1999,
Curtarolo:2000,Ancilotto:2001,Shi:2003,Gatica:2004} 
have been performed on the wetting transition of fluids on various solid 
surfaces.  Most of these studies were focused on simple fluids such as He and 
H$_{2}$ isotopes on alkali metal surfaces. One common feature of these 
systems is that the fluid-surface interaction is only weakly attractive. 
This implies that the bead-up of fluid on the surface at \textit{P}$_{\mathrm{svp}}$ 
will be favorable over the continuous growth of a film. 

Finn and Monson 
\cite{Finn:1989} were among the first to calculate the wetting 
temperature of fluid on solid surface using molecular simulations. 
They predicted the wetting behavior of Ar on solid CO$_{2}$ surface 
using isobaric-isothermal Monte Carlo simulations. %, with 
%\textit{T}$_{\mathrm{w}}^{\ast }$=0.84, \textit{T}$_{\mathrm{pwc}}^{\ast }$=0.94, respectively. 
Shi {\it et al.} \cite{Shi:2002} reevaluated this system using grand 
canonical Monte Carlo (GCMC) simulations plus multiple histogram reweighting 
techniques. Errington \cite{Errington:2004} studied the same system by 
employing a new simulation method.  Bojan {\it et al.} \cite{Bojan:1999} studied wetting behavior of 
Ne on surfaces with various interaction strengths. Curtarolo {\it et al.} 
\cite{Curtarolo:2000} used GCMC simulations to study the wetting behavior 
of inert gases on alkali and Mg surfaces. Shi {\it et al.} \cite{Shi:2003} 
studied the wetting transition of hydrogen isotopes on Rb surface by 
including the quantum effects of the fluids using path integral hybrid Monte 
Carlo simulations. 

 The wetting transition of fluid-fluid systems were studied by 
both experiments and theory, \cite{Bonn:2001} but the wetting behavior of
fluids on solid surfaces was only reported 
for atomistic molecules such as inert gases. No wetting transition has ever 
been seen for any molecular fluids on solids other than hydrogen and its isotopes, 
which are essentially spherical molecules. To our knowledge, the wetting 
transition involving water has not been studied until very recently 
\cite{Gatica:2004} although water is an extensively studied molecule. It 
is known that water does not wet many surfaces (such as graphite) at room 
temperature. Theoretically, the wetting transition of water on graphite is 
expected to occur at a temperature below its bulk critical temperature. In a 
recent paper by Gatica {\it et al.}, \cite{Gatica:2004} wetting temperatures of 
water on graphite had been predicted using a simple model based on the 
water-solid interaction calculated from empirical potentials. The 
recommended \textit{T}$_{\mathrm{w}}$ from their calculations for water on graphite is 
474 K.

In this study, we report evidence for the first-order wetting transition of water 
on graphite from molecular simulations. We estimate the wetting transition 
temperature and 
prewetting critical temperature of water on graphite using grand canonical Monte Carlo 
simulations. The paper is organized as the following: The next section 
describes the potential models and simulation methodology. Section III 
presents the results and discussion. Section IV summarizes our findings. 

\section{POTENTIALS AND METHODS}

Water-water interaction is described by the SPC/E 
model. \cite{Berendsen:1987} This model is widely used in modeling systems 
involving water. The model includes a Lennard-Jones site located on the 
oxygen atom and three partial charge sites on each atom. The parameters for 
this model are given in TABLE \ref{tab1}. The critical temperature 
of water calculated using the SPC/E model is 635 
K. \cite{Hayward:2001} This value is the closest to the experimental value 
(647 K) compared with the predictions by many other popular nonpolarizable 
water potentials. \cite{Guillot:2002} Up to date there is no potential 
available being able to reproduce in every detail the properties of real 
water. \cite{Guillot:2002} It is known that the ability of an interaction 
potential to describe the bulk critical behavior is a necessary (though not 
sufficient) requirement in order for it to predict the wetting transition 
behavior of the fluid on a surface. \cite{Bonn:2001}  Therefore, we chose the SPC/E model 
among tens of water potentials available. 

The graphite surface is modeled as a smooth basal plane. The Lennard-Jones 
interaction between a water molecule and the graphite surface is given by 
the 10-4-3 potential \cite{Steele:1973}

\begin{equation}
\begin{split}
\label{eq1}
V_{\mathrm{sfLJ}} (z)= 2\pi \varepsilon _{\mathrm{sf}} \sigma _{\mathrm{sf}}^2 \Delta \rho _{\mathrm{s}} 
&\bigg{[} 
\frac{2}{5} 
\left( \frac{\sigma _{\mathrm{sf}} }{z} \right)^{10}
-\left( \frac{\sigma _{\mathrm{sf}} }{z} \right)^4	\\
&-\frac{\sigma _{\mathrm{sf}}^4 } {3\Delta(0.61\Delta +z)^3}  
\bigg{]},
\end{split}
\end{equation}
where \textit{z} is the distance between the oxygen atom in a water molecule and the 
graphite surface in the surface normal direction,
 $\sigma _{\mathrm{s}} $ is the number density of carbon atoms in graphite, and $\Delta $ 
is the distance between the graphene sheets in graphite. The graphite 
surface corrugation is not included in this potential.
 The graphite-water interaction parameters 
$\varepsilon _{\mathrm{sf}} $ and $\sigma _{\mathrm{sf}} $ are calculated from the 
Lorentz-Berthelot rules,
\[
\varepsilon _{\mathrm{sf}} =(\varepsilon _{\mathrm{s}} \varepsilon _{\mathrm{f}})^{1/2} ,
\quad
\sigma _{\mathrm{sf}} =(\sigma _{\mathrm{s}} +\sigma _{\mathrm{f}} )/2.
\]
The values of the parameters are: $\rho_{\mathrm{s}}$=114 nm$^{-3}$, $\Delta$=0.335 nm, 
$\varepsilon_{\mathrm{s}}$=0.05569 kcal/mol, and $\sigma_{\mathrm{s}}$=0.340 nm for graphite, 
$\varepsilon_{\mathrm{f}}=\varepsilon_{\mathrm{O}}$=0.1554 kcal/mol, and 
$\sigma_{\mathrm{f}}=\sigma_{\mathrm{O}}$=0.3165 nm for water.

A recently developed effective potential for the dipole-induced dipole, 
dipole-quadrupole, and quadrupole-quadrupole interactions between polar 
fluids and graphite \cite{Zhao:2005} is used to calculate the water-graphite polar 
interactions. For water/graphite system, only 
the induction and dipole-quadrupole terms are important,

\begin{equation}
\begin{split}
\label{eq2}
V_{\mathrm{polar}}(z)=&-\frac{\pi\Delta\rho_{\mathrm{s}}\mu_{\mathrm{f}}^2}{(4\pi\varepsilon_0)^2}    
\bigg{[}
\frac{\alpha_{\mathrm{C}}}{2}
\left(\frac{1}{z^4}+\frac{1}{3\Delta(\Delta+z)^3}\right)  \\ 
&\,\,\,\,\,\,\, +\frac{\Theta_{\mathrm{C}}^2}{3k_{\mathrm{B}}T}
\left(\frac{1}{z^6}+\frac{1}{5\Delta(\Delta+z)^5}\right)
\bigg{]},
\end{split}
\end{equation}
where $\varepsilon_0$ is the vacuum permittivity, $k_{\mathrm{B}}$ is 
the Boltzmann's constant, $\mu_{\mathrm{f}}$ is the dipole moment of the water molecule, 
$\alpha_{\mathrm{C}}$ is the isotropic polarizability of a carbon atom in 
graphite, and $\Theta_{\mathrm{C}}$ is the permanent quadrupole moment on each 
carbon atom in graphite. The values for these parameters are $\mu_{\mathrm{f}}$=1.85 Debye, 
$\alpha_{\mathrm{C}}=1.76\times10^{-3}$ nm$^{3}$, \cite{Miller} and 
$\Theta_{\mathrm{C}}=-3.03\times10^{-40}$C m$^{2}$. \cite{Whitehouse:1909}
We note that $\mu_{\mathrm{f}}$ calculated for SPC/E model is 2.07 Debye, but 1.85 Debye
is chosen here so that comparison can be made between this study and Gatica {\it et al.}'s prediction.

Ewald summations were applied in simulations to account 
for the long-range correction to electrostatic interactions. Since only two 
dimensional periodic boundary conditions along \textit{x} and \textit{y} 
directions were applied in the adsorption simulations, a pseudo-two-dimensional 
Ewald summation method \cite{Yeh:1999} was used. The total electrostatic 
energy is given by 
\begin{equation}
\begin{split}
\label{eq2a}
V_{el}=&\frac{1}{2V_0\varepsilon_0}
\sum_{k{\neq}0}^{\infty} \frac{e^{-k^2/{4\alpha^2}}}{k^2}
\bigg{\vert} \sum_j^N {q_j e^{-i\bf{k} {\cdot} {\bf{r}}_j} } {\bigg{\vert}^2}	\\
&+{ \frac{1}{4\pi\varepsilon_0} } \bigg{[} 
\sum_{n<j}^N { \frac{q_nq_j}{r_{nj}} \mathrm{erfc}(\alpha r_{nj}) }	\\
&-\sum_{mole} {\bigg{(} 
{\frac{\alpha}{\sqrt{\pi}}} {\sum_{j=1}^{site} q_j^2}
+\sum_{n,j}^{list} {\frac {q_n q_j} {r_{nj}} } \bigg{)}}    \bigg{]}
+{\frac{M_z^2}{2V_0\varepsilon_0}},
\end{split}
\end{equation}
where \textit{V}$_{0}$ is the volume of the simulation cell and $\alpha$ is the Ewald convergence 
parameter. In Eq. (\ref{eq2a}), the first term is the reciprocal sum for all the 
Gaussian charges, \textbf{k} is the reciprocal lattice vector, 
\textit{q}$_{j}$ is the partial charge on interacting site \textit{j}, 
\textbf{r}$_{nj}$ is the vector from \textbf{r}$_{n}$ to \textbf{r}$_{j}$. 
The second term is the standard real-space sum, erfc is the complementary 
error function. The third term is the self-exclusion on each molecule, plus 
the 1-2 and 1-3 intramolecular exclusions, where \textit{site} is the total 
number of partial charge sites in the molecule, and \textit{list} contains 
all the 1-2, 1-3 exclusions. The last term is the 2D correction term, where 
\textit{M}$_{z}$ is the \textit{z} component of the total dipole moment of 
the simulation box. 

 \begin{table}%[H] add [H] placement to break table across pages

 \caption{Potential parameters for SPC/E model. \cite{Berendsen:1987} 
$r_{\mathrm{OH}}$ is the O-H bond length, $\theta_{\mathrm{HOH}}$ is the H-O-H bond angle. 
 $\varepsilon_{\mathrm{O}}$ and $\sigma_{\mathrm{O}}$ are Lennard-Jones parameters for
the O atom.   $q_{\mathrm{O}}$ and $q_{\mathrm{H}}$ are the partial charges on O and H atoms, respectively. \label{tab1}}
 \begin{ruledtabular}
 \begin{tabular}{cccccc}
$r_{\mathrm{OH}}$[nm] & $\theta_{\mathrm{HOH}}$ & $\varepsilon_{\mathrm{O}}$[kcal/mol] 
& $\sigma_{\mathrm{O}}$[nm] & $q_{\mathrm{O}}$[$e$] & $q_{\mathrm{H}}$[$e$] \\
\hline
0.1 & 109.47$^{\circ}$ & 0.1554 & 0.3165 & $-0.8476$ & 0.4238 \\
 \end{tabular}
 \end{ruledtabular}

 \end{table}

We have used grand canonical Monte Carlo simulations combined with 
multiple histogram reweighting (MHR) method  
\cite{Ferrenberg:1988,Ferrenberg:1989} to compute the 
saturated coexistence chemical potentials \cite{Shi:2001} and adsorption 
isotherms \cite{Shi:2002} of water on graphite at various 
temperatures.  More details on the MHR method can be found from the original 
literature \cite{Ferrenberg:1988,Ferrenberg:1989} and several 
other articles on its 
applications. \cite{Shi:2001,Pablo:1999}  The GCMC cell 
for adsorption simulations is a rectangular box with volume of 2000$\sigma_{\mathrm{f}}^{3}$. 
The height of adsorption box in the \textit{z} direction, $H$, is 
15$\sigma_{\mathrm{f}}$, the sides in \textit{x} and \textit{y} directions are 
11.5$\sigma_{\mathrm{f}}$. 
The lower plane normal to 
the \textit{z} axis is modeled as the graphite surface. 

Special care was taken to avoid capillary condensation effects on the 
wetting transition properties of the system.
 The plane opposite to the graphite surface is modeled as a hard repulsive wall. The hard wall is
always dry to the liquid phase and wet to the gas phase. 
Therefore it helps to suppress the capillary 
condensation. \cite{Evans:1985,van:1986,Parry:1990,Parry:1992,Finn:1989}
The capillary condensation can also be eliminated by using a simulation cell 
with sufficient separations between the adsorbent surface and the opposite 
hard wall. \cite{Finn:1989,Bojan:1999,Curtarolo:2000,Shi:2003} However, an overall
small cell volume is preferred for the MHR technique. Therefore, we have performed series of
trial simulations using various cell heights to search for an appropriate 
value of $H$. Trial simulations were performed using cell heights ranging from
10$\sigma_{\rm f}$ to 40$\sigma_{\rm f}$ at interested temperatures and chemical
potentials (pressures). It was found that the adsorption properties such as 
isotherms obtained at 15$\sigma_{\rm f}$ are consistent with those obtained
at 20$\sigma_{\rm f}$, 30$\sigma_{\rm f}$, and 40$\sigma_{\rm f}$ within
the statistical fluctuations (see FIG. \ref{isotherm} for an example). This indicates
that 15$\sigma_{\rm f}$ is adequate for effectively avoiding the influence
of capillary condensation for the systems of interest.  Based on this we chose 15$\sigma_{\rm f}$
as the cell height in most of our GCMC simulations. Additional simulations with $H$=20$\sigma_{\rm f}$
were performed as verifications at each temperature. 

The type of move to attempt during a 
GCMC simulation was selected randomly with probability of 0.45, 0.45, 0.05, 
and 0.05 for displacements, rotations, creations, and deletions of a water 
molecule, respectively. Each simulation included equilibration of 80$\times10^6$ 
MC moves and production of 20$\times10^6$ MC moves. Histograms were collected 
every 20 MC moves during the production. The cutoff of pair-wise 
Lennard-Jones interaction between water molecules was 0.9 nm as suggested by 
the original literature, without long-range 
correction applied. \cite{Berendsen:1987}

In the light of Gatica {\it et al.}'s prediction, \cite{Gatica:2004}
we chose to perform GCMC simulations at 460, 470, 480, 490, 500, 510 K with 
varying reduced chemical potentials to obtain the histograms for bulk water 
and water/graphite adsorption systems. The values of bulk saturation 
chemical potentials ($\mu_{\mathrm{svp}}^*$) at each different temperature can be determined using 
the MHR method by combining the histograms collected.
The MHR provide very precise values of \textit{$\mu $}$_{\mathrm{svp}}$ 
through the equal area criterion, \cite{Shi:2001} which is very 
important for studying wetting transitions. Sufficient overlap between 
histograms of adjacent state points is necessary in order to use the MHR 
technique. We employed the method proposed by Shi {\it et al.} \cite{Shi:2002} 
to check the overlap of any two adjacent state points. According to this 
method, the grand canonical partition functions extrapolated by histogram 
reweighting method for any two adjacent state points should approximately 
satisfy 
\begin{equation}
\label{eq3}
\frac{\Xi(\mu_i,V,T_i)}{\Xi(\mu_j,V,T_j)}\bigg{\vert}_{\rm HR} 
\times 
\frac{\Xi(\mu_j,V,T_j)}{\Xi(\mu_i,V,T_i)}\bigg{\vert}_{\rm HR} =1\pm\delta,
\end{equation}
where the subscript HR indicates that the partition function in the 
numerator has been extrapolated from the histogram reweighting of the 
state point in the denominator. The recommended value 
for $\delta$ is 0.65 for checking the overlap of two adjacent 
state points. \cite{Shi:2002} Whenever the overlap criterion defined 
in Eq. (\ref{eq3}) is not satisfied by any of two adjacent state points,
additional simulations at state points that
bridge them were performed. 
From preliminary simulations we found that the wetting 
transition of water on graphite occurs approximately between 470 and 480 K and the wetting critical 
temperature is between 500 and 510 K. In order to narrow down the values of 
$T_{\mathrm{w}}$ and $T_{\mathrm{pwc}}$, 
 we performed additional simulations at 475 K and 505 K. 

\section{RESULTS AND DISCUSSIONS}

\begin{figure}

 \includegraphics[width=75mm]{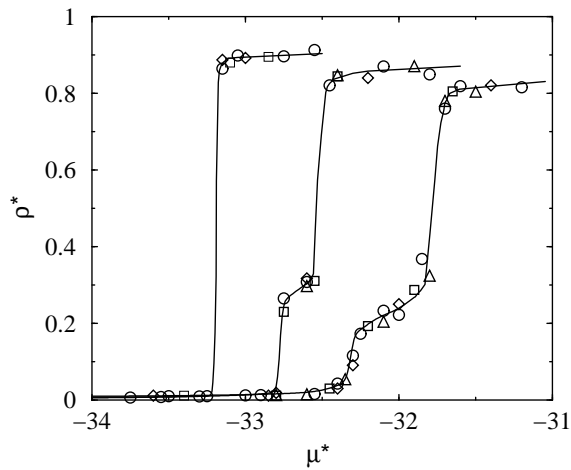}%
 \caption{Typical adsorption isotherms of water on graphite from GCMC simulations 
and MHR, where $\rho^*$ is the reduced number density and $\mu^*$ is the reduced chemical
potential. The curves correspond to $T$=475, 490, 
510 K, from left to right, which were computed from MHR using 
histograms collected from simulations with cell height $H$=15$\sigma_{\rm f}$. The symbols
are data from individual GCMC simulations and were not included 
in the MHR calculations. Circles: $H$=15$\sigma_{\rm f}$, squares: $H$=20$\sigma_{\rm f}$,
diamonds: $H$=30$\sigma_{\rm f}$, triangles: $H$=40$\sigma_{\rm f}$. \label{isotherm}}

 \end{figure}

GCMC adsorption simulations were carried out for reduced chemical potentials up to
 saturation under each selected temperature. Isotherms were calculated 
using MHR based on the histograms collected during the simulations.  Three 
representative isotherms are plotted in FIG. \ref{isotherm}. Some of the data 
points calculated directly from GCMC simulations but not included
in the MHR calculation are also plotted in the 
figure to compare with isotherms from MHR. It can be seen that the difference between
the densities obtained from direction GCMC simulations and those from MHR 
are small. This serves as a test of the accuracy of the MHR isotherms.

To present the isotherms obtained at different temperatures in a concise and
clear fashion, we define a parameter as
\[
\chi^*=\exp\left(\frac{\mu^*-\mu_{\mathrm{svp}}^*}{T^*}\right),
\]
where $\mu^*$ is the reduced chemical potential, $\mu_{\mathrm{svp}}^*$ is 
the saturation chemical potential at the reduced temperature 
$T^*$. Plotting the adsorption coverage versus $\chi^*$ 
gives a diagram similar to FIG. \ref{diagram}, which helps one to identify 
the wetting transition points without ambiguity.
The parameter $\chi^*$ is the ratio of the activity to the activity at saturation, with
$\chi^*=1$ corresponding to $\mu=\mu_{\mathrm{svp}}^*$, or $P=P_{\mathrm{svp}}$. 
Additionally, $\chi^*=P/P_{\mathrm{svp}}$
 if ideal behavior is assumed in the bulk vapor phase. However,  
 water vapor cannot be treated as an ideal gas under the interested 
simulation temperatures in this study. For example, the experimental 
compressibility factor of the saturated water vapor is about 0.865 
\cite{Smith:2001} at 500 K.

\begin{figure}
 \includegraphics[width=75mm]{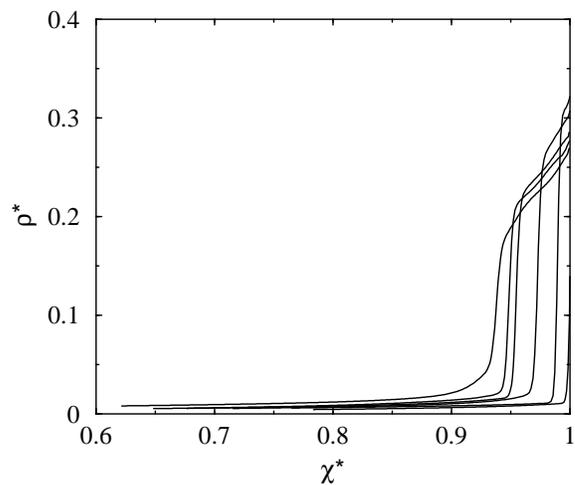}%
 \caption{Adsorption isotherms of water on graphite from GCMC simulations 
and MHR.
The curves correspond to $T$=510, 505, 500, 490, 
480, 475 K, from left to right.\label{fig1}}
 \end{figure}

As the chemical potential was increased 
toward saturation, three different types of behavior in the growth of the water
adsorption film on graphite were observed, corresponding to three ranges 
of temperature. Adsorption isotherms for water/graphite at several 
representative temperatures are shown in FIG. \ref{fig1}. 

\begin{figure}
 \includegraphics[width=75mm]{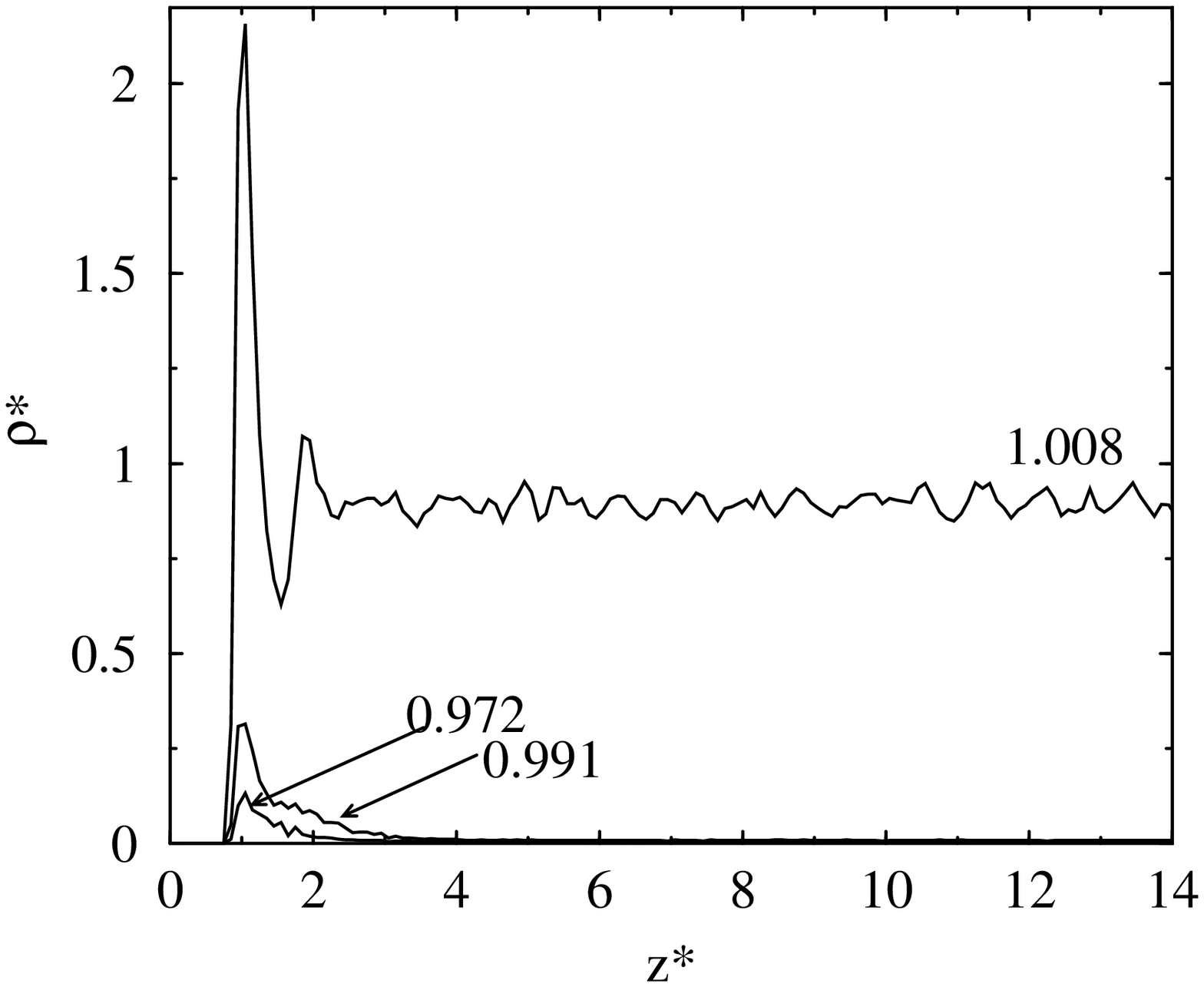}%
 \caption{The local density profiles for water adsorption on graphite as a 
function of reduced distance from the surface, $z^*=z/\sigma_{\mathrm{f}}$, at 475 K.
The values of $\chi^*$ 
 at which the calculations were performed are indicated by the 
labels in the graph.} \label{fig475K}
\end{figure}

At temperatures below 475 K, the adsorption coverage is minuscule until the 
saturation chemical potential is reached, which indicates partial wetting or 
nonwetting. The isotherm jumps to the saturated liquid density at $\chi_{\mathrm{svp}}^* 
$, which can be seen from FIGs. \ref{isotherm}, \ref{fig1}, and from the density profile 
growth patterns shown in FIG. \ref{fig475K}. The sharp increase of density between 
$\chi^*$=0.991 and $\chi^*$=1.008 corresponds to a 
first-order transition from nonwetting to liquid condensation (FIG. \ref{fig475K}). At $\chi^*$=1.008, 
much of the density profile becomes comparable with the 
saturated liquid density profile (the saturated liquid density at 475 K is 
$\rho^*\approx$0.89 from bulk GCMC simulations) except for the first peak 
corresponds to the density of liquid film in contact with the wall. 

\begin{figure}
\includegraphics[width=75mm]{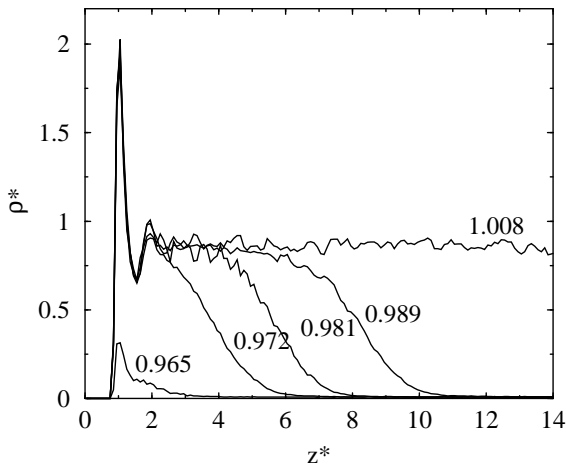}
\caption{The local density profiles for water on graphite as a function
of reduced distance from the surface at 490 K.
The values of $\chi^*$ at which the calculations were performed are indicated
by the labels in the graph.} \label{fig490K}
\end{figure}

In contrast, the simulation results at temperatures between 480 K and 505 K 
manifest quite different behavior. Taking the isotherm at 490 K as an 
example, there is sudden jump in adsorption from minimum to a finite 
coverage of about $\rho^*$=0.25 at $\chi^*$=0.965 to $\chi^*$=0.972 (FIG. \ref{fig490K}). But 
apparently the increased coverage does not correspond to a liquid 
condensation (the saturated liquid density at 490 K is $\rho^*\approx$0.85). 
As the chemical potential is increased further, the film thickens, which 
indicates a wetting behavior (FIGs. \ref{isotherm}, \ref{fig1}, \ref{fig490K}). By comparing the results shown in FIG. \ref{fig1}
 and FIG. \ref{diagram} we readily see that wetting transition of water on graphite occurs 
somewhere between 475 K and 480 K, i. e., $T_{\mathrm{w}} 
$=475-480 K.  At temperatures 
in the range of 480 K and 505 K, the prewetting jump in density occurs 
further to the saturation chemical potential with the smaller density jump 
as temperature increases (FIG. \ref{fig1}). 

\begin{figure}
\includegraphics[width=75mm]{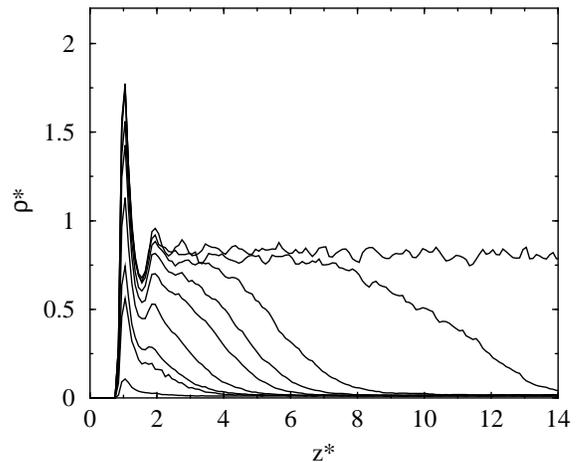}
\caption{The local density profiles for water on graphite as a function of
reduced distance from the surface at 510 K.
Profile curves are for $\chi^*=$
0.731, 0.878, 0.892, 0.919, 0.940, 0.962, 0.977, 0.995, 1.003, 
from bottom to top.\label{fig510K}}
 \end{figure}

At $T$=510 K, the adsorption isotherm becomes continuous as 
the chemical potential increases (FIG. \ref{fig1}), which indicates $T_{\mathrm{pwc}}\le$510 K. 
This is more clearly presented by the growth of density 
profiles shown in FIG. \ref{fig510K}. The adsorption film builds from a thin to a 
thick one continuously with the increase of the chemical potential. The 
density of the first peak in the profiles increases gradually to that of an 
adsorbed liquid. At saturation chemical potential, the density profile 
evolves to the one corresponding to the liquid density except for the first peak 
adjacent to the wall. Comparison of the isotherms obtained at 505 K and 510 K in FIG. \ref{fig1} with FIG. \ref{diagram}
indicates that the prewetting critical temperature of water on graphite lies  
somewhere between these two values of temperature, i.e. $T_{\mathrm{pwc}}$=505-510 K. 

\begin{figure}
 \includegraphics[width=75mm]{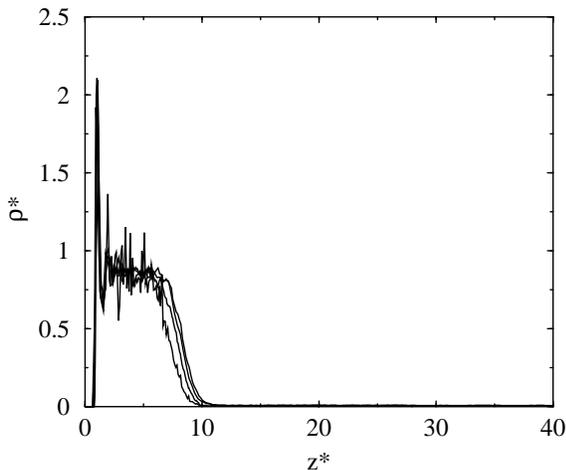}%
 \caption{Film density for water on graphite at \textit{T}=490 K and  
$\chi^*=$0.992. Curves correspond to varying heights of the simulation 
cell: $H^*$=10, 20, 40, 30, from left to right. 
\label{fig3}}
 \end{figure}

The nature of the prewetting jump of water on graphite at temperatures 480-
505 K can be further shown by the results obtained from simulations at 490 K, 
$\chi^*=$0.992, with varying simulation cell dimension in the surface 
normal direction. Shown in FIG. \ref{fig3} are the density profiles obtained from 
simulations with cell heights of $H^*=h/\sigma_{\mathrm{f}}$=10, 20, 30, 40, respectively. 
In order to compare the results with consistency, in those four simulations 
the area of the graphite wall is kept at 10$\sigma_{\mathrm{f}}$$\times$10$\sigma_{\mathrm{f}}$, 
but the height, or the volume, of the cell varies.
It can be seen from FIG. \ref{fig3} that the rapid 
rise of the film thickness to a finite value is independent of the height of 
the simulation cell. The film thickness keeps at about 7.5$\sigma_{\mathrm{f}}$
 under various $H^*$. This is a clear 
indication that the transition is prewetting rather than capillary 
condensation. We also notice that the density profile at $H^*$=10 
has more fluctuations compared with those at $H^*$=20, 30, and 40. 
That could be due to the finite size effects in 
$H^*$. This confirms the importance of using a simulation cell 
with sufficient height in order to obtain reliable wetting transition 
information. Theoretically the density profiles obtained with different 
$H^*$ should coincide with each other. But it is hard to 
achieve this in simulations because of statistical fluctuations and 
metastability nature of the problem. 

One of the clearest demonstrations of the first-order nature of many wetting 
transitions is the observation of one stable and one metastable states  
shown in the systems of interest in the temperature 
range of \textit{T}$_{\mathrm{w}}$ to \textit{T}$_{\mathrm{pwc}}$ \cite{Bonn:2001}. The 
experimental work by Bonn {\it et al.} \cite{Bonn:1975} showed that two different 
stable values of the film thickness could be found 
in the binary liquid mixture of methanol/cyclohexane at a temperature between 295 K and 
308 K. The thin film of 1.0 nm is the metastable state, and the thick 
film of 40.0 nm is the stable state. Shi {\it et al.} \cite{Shi:2003} 
observed the similar switching behavior between thin and thick films in the 
simulation study of hydrogen isotopes on alkali metal surfaces. 

\begin{figure}
 \includegraphics[width=80mm]{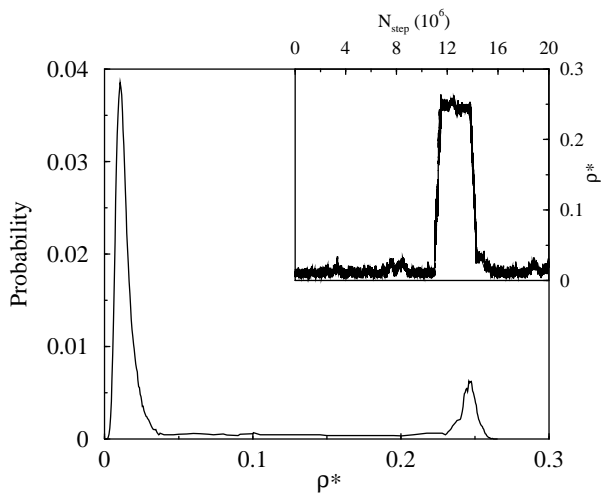}%
 \caption{The switching between the thin and thick film of water adsorption 
on graphite at \textit{T}=490 K, $\chi^*=$0.968. The inset shows 
the density evolution during the simulation. \label{fig4}}
\end{figure}

We also observed the coexistence of stable and metastable states for water 
on graphite, which is shown in FIG. \ref{fig4}. The probability density function of 
the water density distribution in the box is collected during the simulation 
with \textit{T}=490 K, $\chi^*=$0.968. One major peak is located at $\rho^*\approx$0.012 
corresponding to a thin film, at which the simulation 
samples most frequently. Another small peak exists at $\rho^*\approx$0.25, 
which corresponds to a thick film. The bimodal 
feature of the probability density function of the distribution indicates 
the wetting transition at this temperature is first order. The switching between
the thin and thick films is further confirmed by the fluctuation of the number 
density of adsorbate in the box. In the inset of FIG. \ref{fig4} we show the 
evolution of water density in the simulation cell as a function of 
simulation steps. At about $11\times10^6$ configurations the system 
abruptly jumps to a higher density of $\rho^*\approx$0.25 from 
$\rho^*\approx$0.012, corresponding to a switching from thin film to 
thick film. This thick film subsequently evaporates to the thin film at 
about $14\times10^6$ configurations. Comparing the sizes of the two 
peaks and evolution of the number density fluctuation we readily conclude 
that at this state point, the thin film is stable and the thick film 
was metastable. 

In this work we did not attempt to determine the exact values of
the wetting temperature or the prewetting critical temperature, but only provide estimated ranges for them, 
which are 475-480 K and 
505-510 K, respectively. One reason is the lack of reliable theoretical method for determination of exact 
$T_{\mathrm{w}}$ and $T_{\mathrm{pwc}}$. 
One of the popular techniques being used previously is 
 a power law extrapolation method \cite{Ancilotto:1999}. 
Theoretical predictions indicate
that $\Delta\mu^*=(\mu_{\mathrm{svp}}^*-\mu_{\mathrm{w}}^*) \propto (T^*-T_{\mathrm{w}}^*)^{3/2}$ \cite{Ancilotto:1999}.
Hence, a plot of $\Delta\mu^*$ versus $T^*$ can be used to identify 
$T_{\mathrm{w}}^*$ by extrapolating the curve to $\Delta\mu^*=0$ \cite{Mistura:1994}, with
the saturation and wetting transition chemical potentials 
($\mu_{\mathrm{svp}}^*$ and $\mu_{\mathrm{w}}^*$) at various temperatures
up to $T_{\mathrm{c}}$ of the fluid being calculated from MHR. 
However, Shi {\it et al.} \cite{Shi:2002} found that this power law extrapolation method
could be quite inaccurate in predicting the wetting temperature of certain system. 

Another important concern is the realism of the water potential employed in this study.
It is known that the transition temperature calculated from simulations is sensitively dependent on  
the solid-fluid interactions \cite{Curtarolo:2000,Shi:2003}. For example, Shi {\it et al.} 
found that a $\sim $10{\%} increase in the surface-fluid attraction 
decreases the wetting transition temperature of Ar on a CO$_{2}$ surface by 3 K \cite{Shi:2003}. 
In this work, the choice of the 
water potential will affect the graphite-water interaction implicitly, and thus the calculated $T_{\rm w}$.
The SPC/E water potential gives by far the best predictions for the critical properties
of the bulk water among all the available nonpolarizable water models,
while it still cannot simulate exactly the coexistence properties of
real water. The accuracy of the estimated $T_{\rm w}$ and $T_{\rm pwc}$ may be improved
by using more accurate polarizable models such as the Gaussian charge polarizable model \cite{GCPM}.

In addition, by modeling the graphite as a smooth surface, we neglected
the possible impact from surface corrugations and dynamics of the surface structure during the adsorption.
A previous simulation work indicates that impact of surface corrugation of the adsorbent
on the wetting transition behavior of Ne is minimal \cite{Bojan:1999}. But it is unclear
if the same conclusion is applicable to the graphite-water system.

Cheng {\it et al.} \cite{Cheng:1993} proposed a simple model (CCST hereafter) which 
interprets the wetting transitions in terms of a balance between the surface 
tension cost of producing a thicker film and the energy gain associated with 
the film's interaction with the surface, \textit{V}(\textit{z}). The theory 
results in an implicit equation for the wetting temperature 
\begin{equation}
\label{eq4}
I=-\int_{z_{\mathrm{min}}}^{\infty} V(z)dz=\left( \frac{2\gamma}{\rho_l-\rho_v} \right)_{T_{\mathrm{w}}},
\end{equation}
where $\rho_l$ and $\rho_v$ are the number densities 
of the adsorbate liquid and vapor at coexistence, $\gamma$ is the 
surface tension of the liquid, and $z_{\mathrm{min}}$ minimizes the 
fluid-surface interaction potential \textit{V}(\textit{z}). The wetting transition
temperature can be calculated by solving the equation since the right hand 
side of Eq. (\ref{eq4}) is dependent implicitly on temperature. The impact of solid-fluid
interaction on $T_{w}$ is reflected by $I$, and the fluid-fluid interaction is incorporated in the model
by $\gamma$ and $\rho_l-\rho_v$. 

The wetting transition of water on graphite calculated in this work agree 
quantitatively with the previous prediction \cite{Gatica:2004} using the 
CCST model, although different water potentials are employed in the two 
studies. It has been pointed out by Shi {\it et al.} \cite{Shi:2003} that 
the wetting behavior predicted theoretically depends on both the well depth and well shape of the solid-fluid 
interacting potential. 
Here the potential width is defined as the full width 
at half minimum of the attractive part of the potential. 
In Gatica {\it et al.}'s work, TIP4P 
was used instead of SPC/E. But we note that the well depth (\textit{D}) and well 
width (\textit{w}) of the water/graphite potential for SPC/E and TIP4P are 
almost identical, both with $D=$9.35 kJ/mol and 
\textit{w}=0.135 nm, if evaluated at \textit{T}=475 K and using the water dipole 
moment of 1.85 D. 
Therefore we expect 
that the wetting transition temperature calculated from the CCST model using 
these two potentials be comparable. If the simulation results in this work 
are taken to be standard, the CCST model predicted $T_{\mathrm{w}}$ of 474 K is very 
accurate indeed. It has been shown that the CCST model usually works well in 
predicting the wetting behavior involving spherical fluids such as inert 
gases, but it has not been tested extensively with nonspherical molecules. The 
fact that this simple model works well in predicting the wetting of 
water on graphite, although water has a very different kind of potential 
than inert gases, indicates that the CCST model contains the essential physics of 
wetting.

\section{CONCLUSIONS}

In summary, we report the first simulation study of the wetting 
transition of water on graphite surface. The wetting transition temperature 
calculated from GCMC simulations is 475-480 K, and the prewetting
critical temperature is 505-510 K. The wetting transition is first 
order. The simulation results in this work
agrees well with the prediction by the CCST model, although the CCST model is 
designed on the basis of the simple fluids such as inert gases. 

Finally, we point out that the wetting temperature and prewetting critical temperature
calculated in this work depends on the accuracy of the water potential employed. 
Improvement in the predictions may be made if more accurate water potential
is available. Future investigations can be performed by including
the corrugation of graphite surface, the finite size effect of the system, and by 
using the more robust simulation techniques such as the one proposed by Errington \cite{Errington:2004}. 
 Experimental search for the predicted wetting behavior is also warranted.

\begin{acknowledgments}
The author thanks Peter T. Cummings and Milton W. Cole 
for many helpful discussions throughout this work. This research was 
conducted at the Center for Nanophase Materials Sciences, which is sponsored 
at Oak Ridge National Laboratory by the Division of Scientific User 
Facilities, U.S. Department of Energy.
\end{acknowledgments}

\bibliography{wetting}

\end{document}